# Prominence-cavity regions observed using SWAP 174A filtergrams and simultaneous eclipse flash spectra

by:

C. Bazin[1] , S. Koutchmy[1] and E. Tavabi[2]

1- Institut d'Astrophysique de Paris, UMR 7095, CNRS & UPMC

2- Physics Department, Payame Noor University, 19395-3697 Teheran   I.R. of Iran

**Abstract**: Images from the SWAP (Proba 2 mission) taken at 174A in the Fe IX/X lines are compared to simultaneous slitless flash spectra taken during the last solar total eclipse of July, 11$^{th}$ 2010. Many faint low excitation emission lines together with the HeI and HeII Paschen Alpha chromospheric lines are recorded on eclipse spectra where regions of limb prominences are obtained with space-borne imagers. We consider a deep flash spectrum obtained by summing 80 individual spectra to show the intensity modulations of the continuum. Intensity depressions are observed around the prominences in both eclipse and SWAP images. The prominence cavities are interpreted as a relative depression of plasma density, produced inside the corona surrounding the prominences. Photometric measurements are shown at different scales and different, spectrally narrow, intervals for both the prominences and the coronal background.

**Key words**: Prominences; Prominence- Corona Interface; Helium spectra

## I) Introduction

Prominences and their disk counterpart, filaments, rise to coronal height and are cool objects formed above a polarity inversion line (PIL) of the large scale photospheric magnetic field. Their thermodynamical properties are well described by a standard model (Engvold et al. 1990) showing a large dispersion of the critical parameters; the latest developments of both high resolution groundbased and space observations still show several open issues (Labrosse et al. 2010). White-light (W-L) photographic eclipse images have shown features that have been called "cavity separations", in projection on the plane of the sky as well as, the prominence and the overlying system of arches, with the streamer sheet extending radially to several radii (see Saito and Hyder, 1968, Saito and Tandberg-Hanssen 1973, Tandberg-Hanssen, 1995). Sometimes the filament/prominence/cavity system erupts, so that the analysis of these regions, including the origin of the heating presumably responsible for the eruption or accompanying it, is of great interest.

Broad band W-L eclipse CCD or CMOS images, taken in different wavelengths, now also show the limb prominences (November and Koutchmy, 1996, Koutchmy, Filippov and Lamy, 2007, Jejcic and Heinzel, 2009), allowing a more quantitative analysis. The use of eclipse slitless spectra, taken with a modern detector makes the spectral analysis possible, including the measurement of properly filtered W-L radiations.

EUV (Extreme Ultra violet) filtergrams, obtained in space well before and after the eclipse, permit a more extended temporal analysis of these regions, including the dynamical effects. and the 3 D geometry, taking advantage of the changes of geometry, due to the solar rotation. This allows the description of the behavior of the corona around prominences and the structure of the magnetic field in the context of the filament channel, the PIL and the streamer association. However, it is clear that the lack of spatio-temporal resolution at the scale of interest for the heating (of order of 100 km or less and few seconds of time or less) is a limitation that should be kept in mind when the dynamical phenomena are discussed, in addition to the lack of relevant direct magnetic field measurements in these coronal regions.

The total eclipse of $11^{th}$ July 2010 (Pasachoff et al. 2011 and Habbal et al. 2010- 2011 for a description of the corona at this eclipse), allowed us to observe the true continuum between helium prominence lines (He I 4713A and Pα He II 4686A), taking into account faint low excitation emission lines using fast slitless spectra taken at the contacts, before and after the totality. These observations also yield relatively new results on the determination of the electron densities inside prominences when compared to previous photographic ground-based coronagraphic filtergram observations in the D3 line, where the continuum, at the location of the prominences, was not precisely measured (see Kubota and Leroy, 1970 for Lyot-coronagraph observations and Koutchmy, Lebecq and Stellmacher 1983 for eclipse broadband observations). On flash spectra, measuring the continuum outside prominences, allows us to study the electron density of the cavity and compare with the Proba 2/SWAP (Sun Watcher with APS detectors and image Processing) EUV filtergrams, dominantly recording the Fe IX/X coronal lines (Sirk et al. 2010). SWAP is the main instrument of the Proba 2 mission (Project for OnBord Autonomy), see Berghmans et al. 2006, Defise et al. 2007, and Halain et al. 2010 for a description of the SWAP parameters (see De Groof et al 2008a and **2008b** for the CMOS-APS imaging detectors). Proba-2 is an ESA microsatellite launched in 2009. The SWAP imager has a large field of view of 54 x 45 arc minutes compared to the field of view of the well-known EIT imager of SoHO, and has a typical cadence of 1 image per min.

The presence of many faint low excitation emission lines in our flash spectra[1], such as lines of Ti +, Fe +, Mg+ etc. could eventually reveal the importance of the First Ionisation Potential (FIP) effect which occurs in the low layers of the Transition Region (TR), but their discussion is beyond the scope of this paper. Note that these faint emission lines, immediately outside the limb of the Sun, are impossible to observe properly outside total solar eclipses, due to the effect of parasitic scattered light of instrumental origin, even with a Lyot coronagraph which usually over-occults the Sun.

---

[1] From the historical point of view, flash spectra were rather long exposure frames using photographic plates at the exit of a slitless spectrograph. They were taken just after the last Baily's beads seen in white light, when the chromospheric lines are still faint and are superposed on the myriad of narrow low excitation emission lines overlying the very faint photospheric continuum left outside the solar limb. They are revealed at the coronal intensity level during total eclipse, when the level of parasitic scattered light is completely negligible.

A mapping of the intensity ratio of HeI/HeII lines (at 4713A and 4686A) inside the SWAP cavity images will also be used for the study of its distribution, as well as for the temperature evaluation of the corona-prominence region, where rather surprising coronagraphic temperature measurements were reported in the past, based on these lines and their profiles (Hirayama and Nakagomi, 1974 and Hirayama and Irie, 1984). Here we have the advantage of having many W-L images of the corona taken during the eclipse totality (Pasachoff et al 2011, Habbal et al. 2011, Koutchmy et al. 2011).

**II) Observations.**

1) **Global view of the corona during the 11$^{th}$ July 2011 total eclipse.**

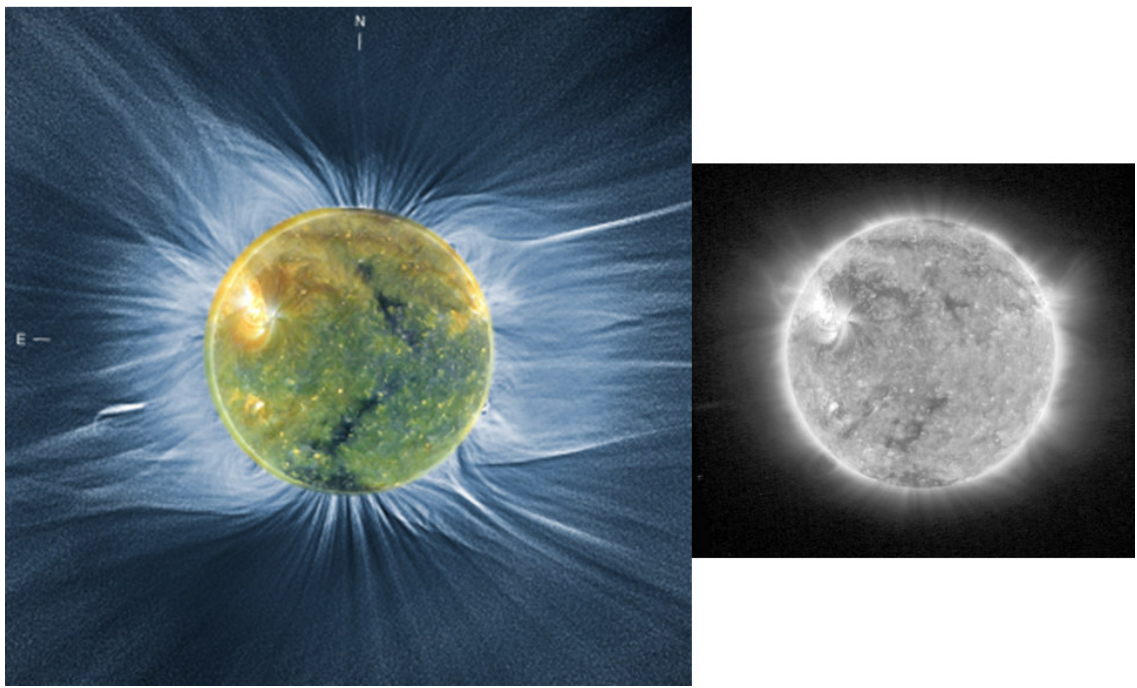

1-a                                            1-b

Figure 1a *: Composite of a processed W-L total eclipse image of the corona on which is superposed an AIA 194A image made almost simultaneously, with proper scaling and orientation. 1b: Quasi-simultaneous SWAP stacked filtergram at 174A, corresponding to the totality of the 11th July 2011 eclipse observed from Hao (French Polynesia), at the same scale.*

The ground based experiments carried out in the French Polynesia produced several images taken in W-L. They allow identification of the streamers and arches (see Figure 1a) with an image taken at 18h43 UT and a SWAP stacked EUV filtergrams taken near 19h UT (Figure 1b). From the Figure 1a and 1b images, arches above limb prominences were identified, especially the one at the S-E limb, where a cavity is seen in W-L with the corresponding prominence inside, also observed in the helium lines using our flash spectra. The cavities are dark and surround the prominences close to the limb. More data were obtained thanks to the high cadence flash spectra, which allow a precise evaluation of the heights above the lunar profile. The emission lines could then be measured from the resulting light flux, seen above the edge of the Moon, at the low heights where the magnetic field is known to dominate the kinetic pressure.

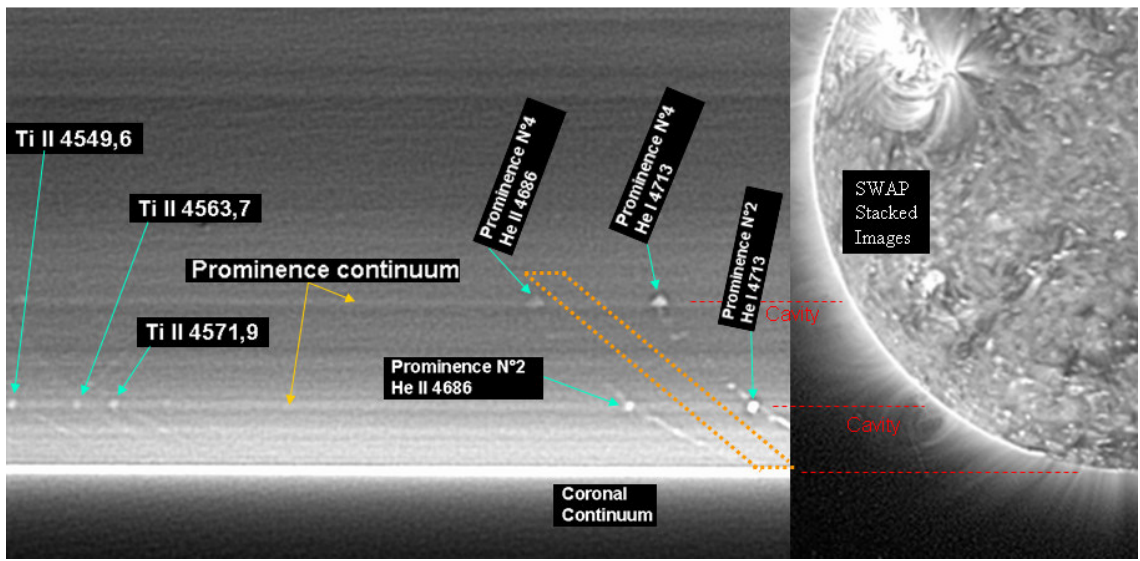

Figure 2: *Processed partial image from the flash spectra sequence, showing the continuum between helium lines prominence emissions. 8s of integration was used (from images taken near the 2$^{nd}$ contact at Hao, from 18h41:53,30s to 18h42:0,14s), to show the "cavity" effect (decreased fluxes). The enhancement due to the prominences seen in the true continuum (left), is compared to a partial SWAP processed image, taken in the same location at South-East limb, (right). The orientation and scale of the SWAP image are the same as in the spectrum. The dashed orange lines show the region used to perform the photometric analysis of the continuum which is plotted on figures 4a and 4b. It is situated between the HeI and HeII emission lines. A scheme of the experiment is given in Appendix (see also Bazin et al. 2011).*

We also performed photometric cuts outside the helium emissions, in the continuum, along the direction shown with dashed orange lines in Figure 2, to deduce the intensity variations in the prominence-corona region. Figure 2 shows the comparison of a composite made of 80 stacked eclipse flash-spectra with the SWAP filtergram taken at the time of the eclipse totality; the location where the photometric cuts were performed is indicated.

**2) Analysis of the Prominence - Corona region using SWAP, AIA and flash spectra in the region of the 4686A He II line.**

The cavity looks relatively dark in W-L due to a deficiency of electrons or plasma density (Saito and Tandberg-Hanssen, 1973). A naïve interpretation of this long duration phenomenon would be to consider that the missing mass fills the prominence magnetic structure. This question is still open and linked to the so-called prominence-corona interface (PCI) physics, to small scale plasma flows around the prominence and to the topology of the magnetic field (Hirayama, 1964). It could be the result of an intermittent condensation process of suddenly, radiatively cooled coronal gas, falling towards the complex prominence magnetic structure imbedded into the surrounding corona. This process occurs at very small scales (see Koutchmy, Filippov and Lamy, 2007), where turbulent motions are usually assumed. New eclipse observations should now be compared with space-borne EUV filtergrams obtained simultaneously, in order to further investigate the former idea. The following images were obtained by AIA (Atmospheric Imaging Assembly) of SDO (Solar Dynamic Observatory), archived by NASA- Goddard Space Flight Center. AIA of SDO takes one image every 12 second in each wavelength of coronal lines (see the web site of AIA/SDO) with a considerably improved spatial resolution but a field of view somewhat smaller than that of SWAP. Prominences are seen in emission in the HeII 304A resonance line on AIA filtergrams and, simultaneously, in the ionized helium P$\alpha$ line of HeII at 4686A, using our flash spectra obtained during the total eclipse**.** Note that the P$\alpha$ 4686A line is considerably less optically thick than the 304A resonance line. (see e.g. Labrosse et al 2010).

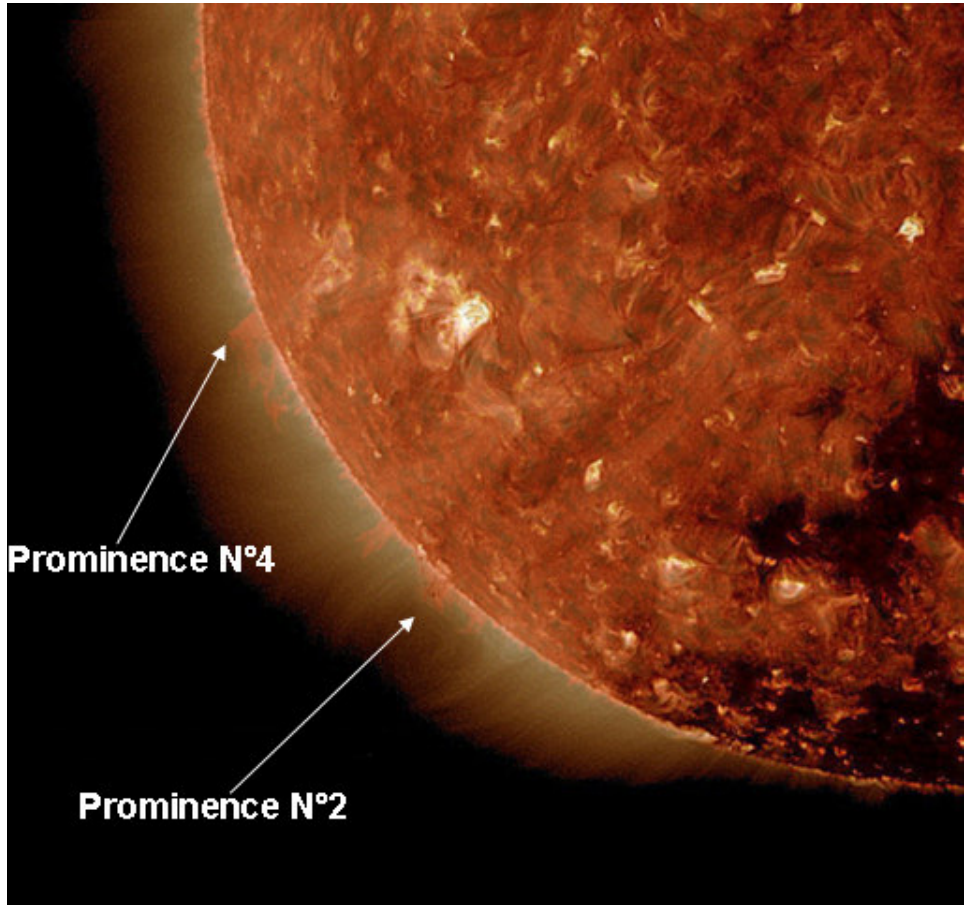

*Figure 3 Superposition of the 304A and the 193A emissions in the partial frame image at SE taken at the time of the eclipse using AIA(SDO). The arrows points to the studied prominences.*

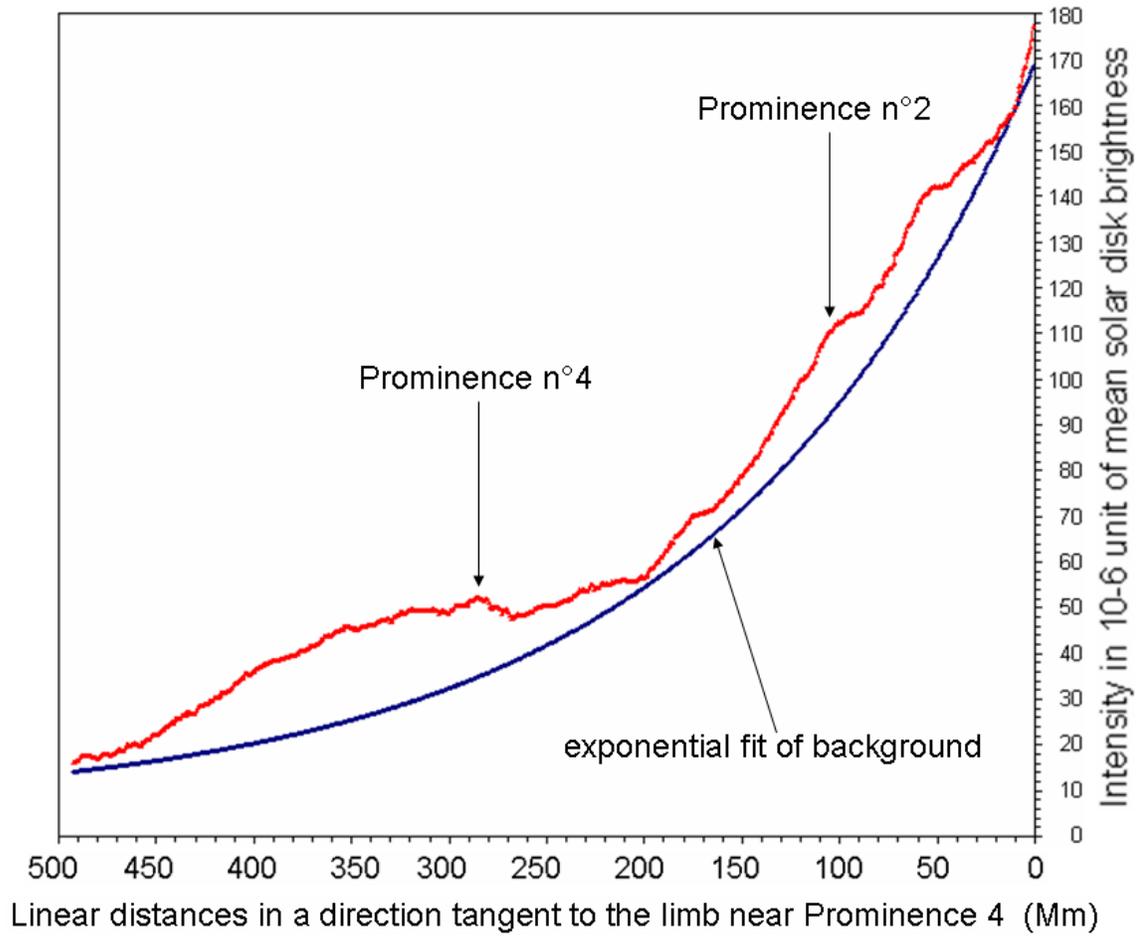

Figure 4a: *Photometric cut through the flash spectrum corresponding to the W-L "clean" continuum dashed region shown on Fig 2. The blue curve represents the background that we subtracted to deduce figure 4b.*

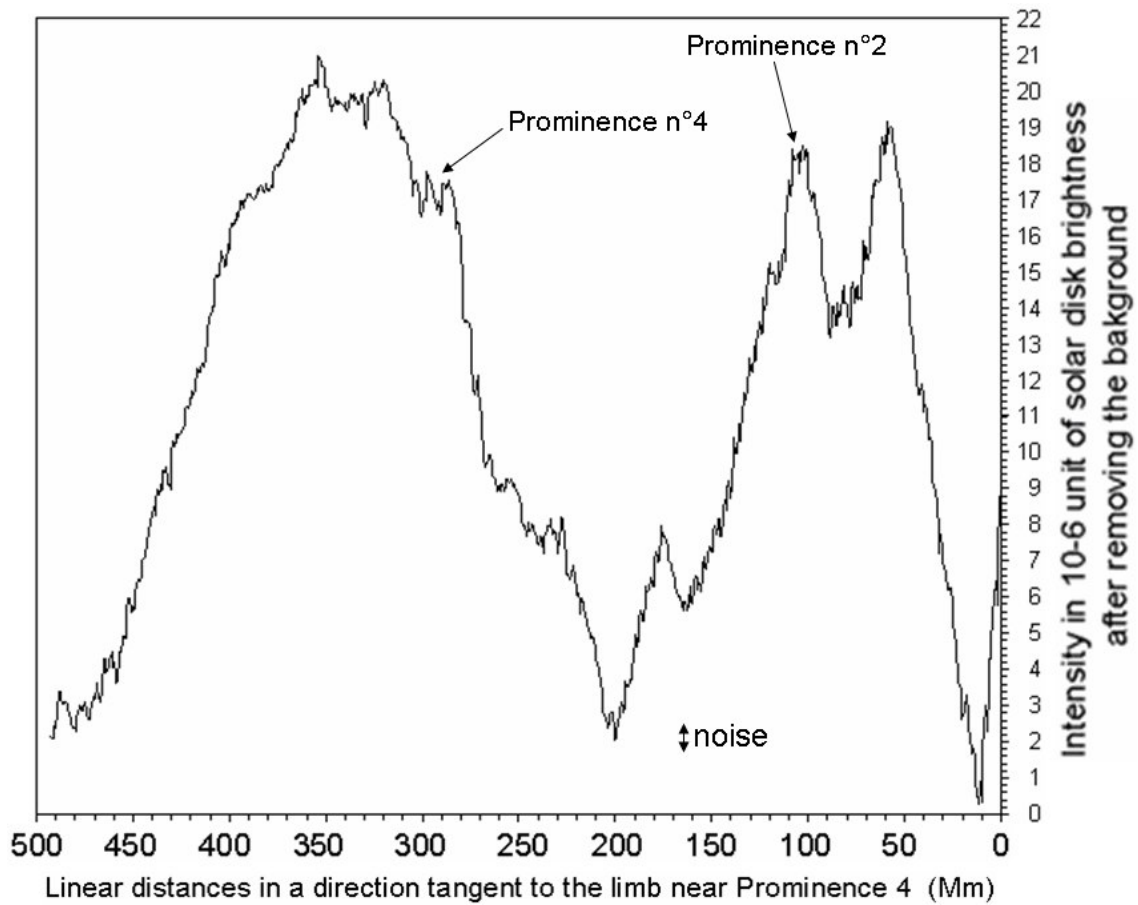

*Figure 4b- Photometric cut deduced from Figure 4a, after removing the background corresponding to the shown fitting in the continuum.*

The graph of Figure 4a shows the result of superposing original photometric cuts taken linearly along the direction tangential to the limb (see the dashed lines in Figure 2), and not exactly along the local limb which is obviously curved. This allows us to increase the signal to noise ratio in the region between the prominences and to distinguish both the true continuum of each prominence superposed on the local coronal continuum recorded along the line of sight and the depressed continuum outside the prominence, corresponding to the cavity parts. This is indeed a difficult task because of the overlapping of structures along the spectral dispersion direction when a slitless spectrum is used. Note that a selection effect similar to the effect produced by the slit of a fictive spectrograph exists just because the radial extension of all recorded emissions is drastically limited by the scale heights of each relatively cool emission line, and, to a smaller extent, the continuum of the chromosphere. The complication comes from the fact that the spectral dispersion is not exactly along the radial direction and, worse, because the direction of dispersion changes with respect to the radial direction. The deviations of the observed intensity plotted in figure 4b deduced from the smoothed fit, correspond to the continuum intensity variations and modulations, coming from prominences seen at higher altitudes and the nearby region. Additional brightenings in HeI 4713A and He II 4686A emission lines are seen as shells (see Bazin et al 2011) bounded on one side by the Moon profile (made of mountains and valleys), but they will not be considered here**.** Accordingly, our quantitative results on prominences alone should be considered with caution due to line of sight integration effect. Some evaluation of the coronal cavity effect, including the prominence-corona region seen in W-L "clean" continuum near the 4686A helium line, has however been attempted (Figures 4a and 4b). The RMS noise level is estimated to be less than $1 \times 10^{-6}$ units of the mean solar disk intensity. The deficiency is evaluated to be $4 \times 10^{-6}$ in units of the mean solar disk brightness in the case of prominence n°2, where the apparent coronal continuum background drastically decreases with radial distance. The deficiency is $3.5 \times 10^{-6}$ for prominence n°4, where the coronal background is less sharply decreasing. This represents a relative value of 7% for prominence 4, which is almost 10 times less than the one we obtained using the SWAP filtergrams at 174A: 60 to 70% (see further Figure 9).

The analysis of the deficiency needs more investigation in order to understand its relationship with the corresponding prominence. This is why we now analyze the intensity ratio of HeI 4713A/ HeII 4686A by making a mapping, where the signal is more significant. We note in Figures 4a and 4b that in the continuum, the amplitude of the signal corresponding to the precise location of the prominences is typically of the same order as the signal of the deficiency, possibly arguing in favor of the idea that the missing gas from the low parts of the cavity is filling the prominence. This is probably the $1^{st}$ time that the true continuum of a prominence is spectroscopically compared to the surrounding coronal background, thanks to the use of this deep flash spectrum obtained after integrating almost 100 spectra. Filtergrams used in the past often contained some contribution from line emissions, which created the impression that electron densities are higher than what they are in reality. Obviously, the contribution of the prominences is well recorded over the flash spectra at the location of the He lines, even in the case of a faint line like the 4686A Pα line (see Figures 2 and 5).

### 3) Mapping of the intensity ratio of HeI 4713A / HeII 4686A

Figure 5 shows the prominences seen in the Helium lines during the eclipse totality and using the optically thick 304A line from AIA/SDO simultaneous filtergrams.

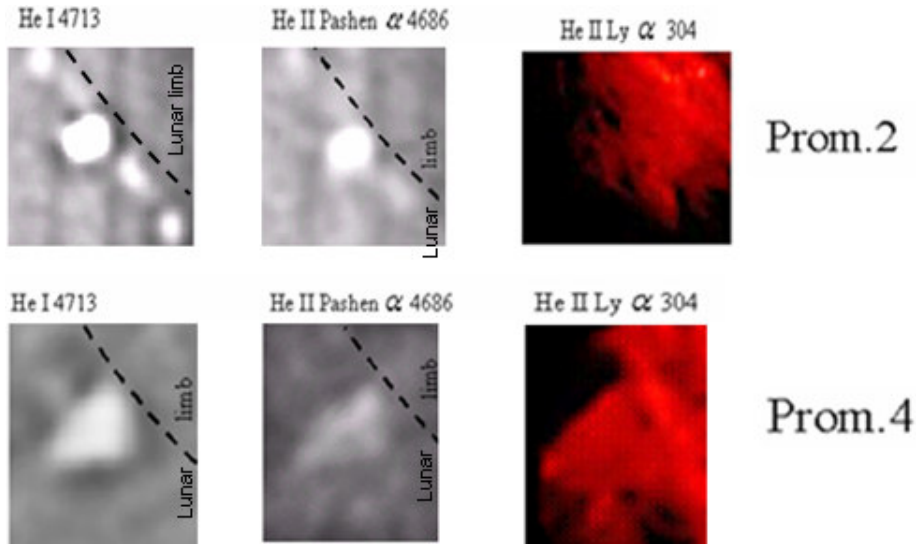

Figure 5: *Images of 2 prominences in the 4713A line of HeI (at left), in the **P α** 4686A line of HeII (in the middle) and in HeII Ly α line at 304A (several order of magnitudes thicker than the Pα line).*

We now consider the ratio of intensities HeI (4713A)/ HeII (4686) which possibly reveals the temperature structures of the prominence 2 (see figure 5) and of the surrounding region.

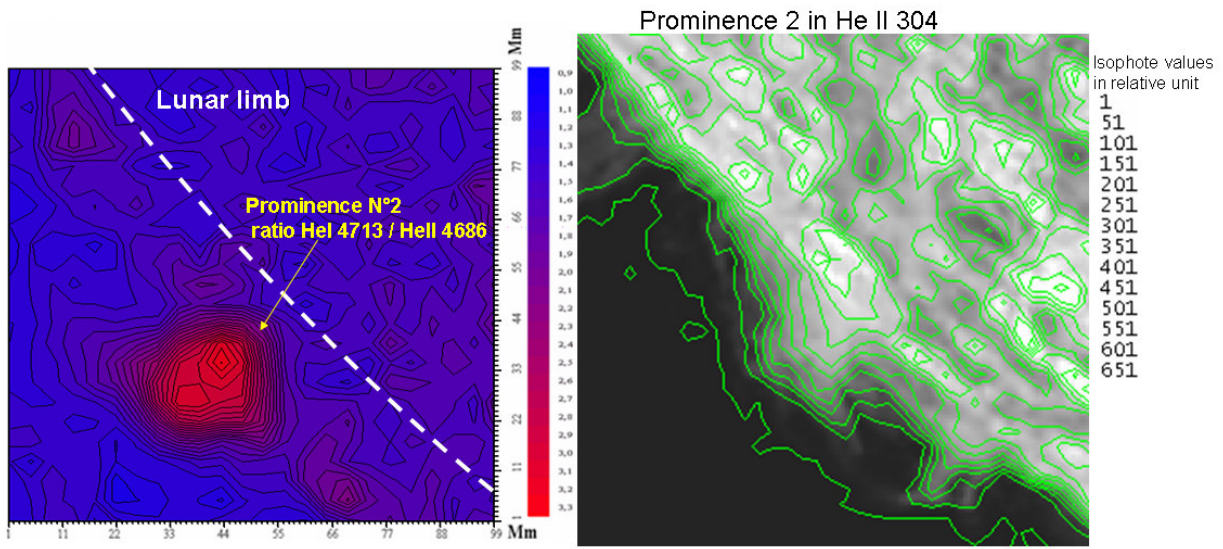

Figure 6- *Left : 2D mapping showing that inside the core of the prominence 2 the relative intensities of the Pα emissions of ionized helium are 3 times smaller than around or otherwise, that this HeII Pα emissions are relatively enhanced in the outer parts of the prominence, compared to the neutral HeI emissions. Right : isophote levels of the same prominence N°2 in the 304A HeII line, observed at 19h19 TU with AIA of SDO.*

The maximum values of the intensity ratio of HeI (4713A)/HeII (4686A) is found to be 3.3 (Figure 6). This is in agreement with the Hirayama and Irie (1984) photographic results and with what was observed in the chromosphere by the same authors, interpreted as a result of the ionization of the helium lines by UV radiations (Hirayama, 1971). The intensity ratio increases abruptly near the prominence edge, where the P-C region is located, by a factor 2 over a distance of 10 Mm. This needs to be compared with the surrounding cavity, observed with SWAP/Proba2 in the Fe IX/X line at 174A (see Figure 2), where the temperature reaches 0.6 to 1 MK (Sirk et al. 2010). Surprisingly, the SWAP image also shows, at small scale, some coronal emission near prominence 2, with the dark cavity around.

**4) Photometric analysis of the cavity regions in W-L, with SWAP and with EIT/SOHO.**

An image taken in W-L at the time of the 11$^{th}$ July 2010 eclipse totality in Hao (French Polynesia) was used. We compare the W-L intensity profiles corresponding to the K-corona with SWAP 174A profiles, in order to analyse the cavity regions. The F-corona has been subtracted, by using the F-corona radial intensity profile from Koutchmy and Lamy (1985). The integration was performed over a 20 Mm width to get a better signal to noise ratio. Tangential and radial intensity cuts are performed around the prominences (from #110 stacked SWAP images converted into polar coordinates).

Figures 7 and 8 show the region of the cavity with the surrounding corona in polar coordinates.

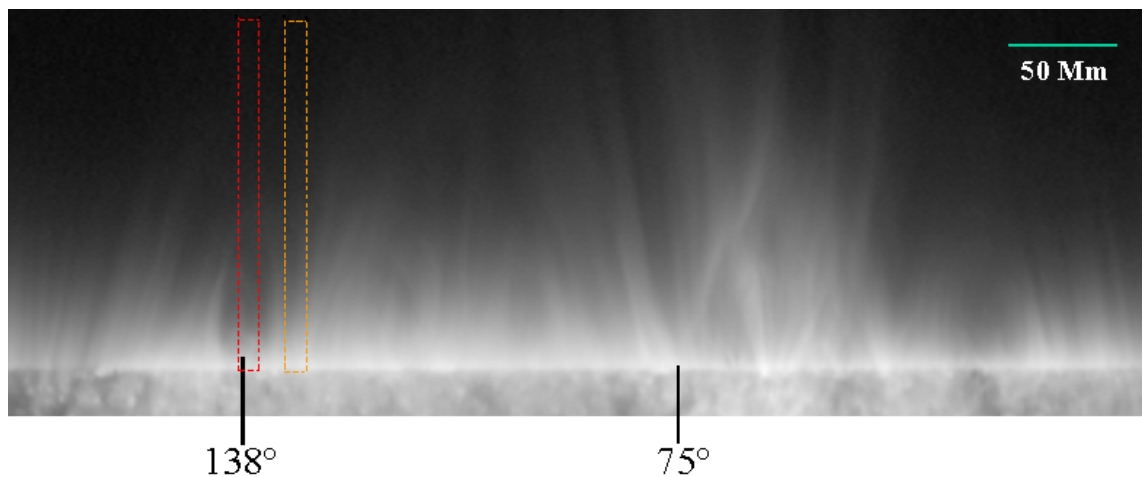

Figure 7: *SWAP image taken at the time of the eclipse, in polar coordinates. The studied cavity of prominence 2 is located at 138° heliocentric coordinates. The red and orange dotted lines show where the radial cuts are taken along and outside the cavity, see Figure 10.*

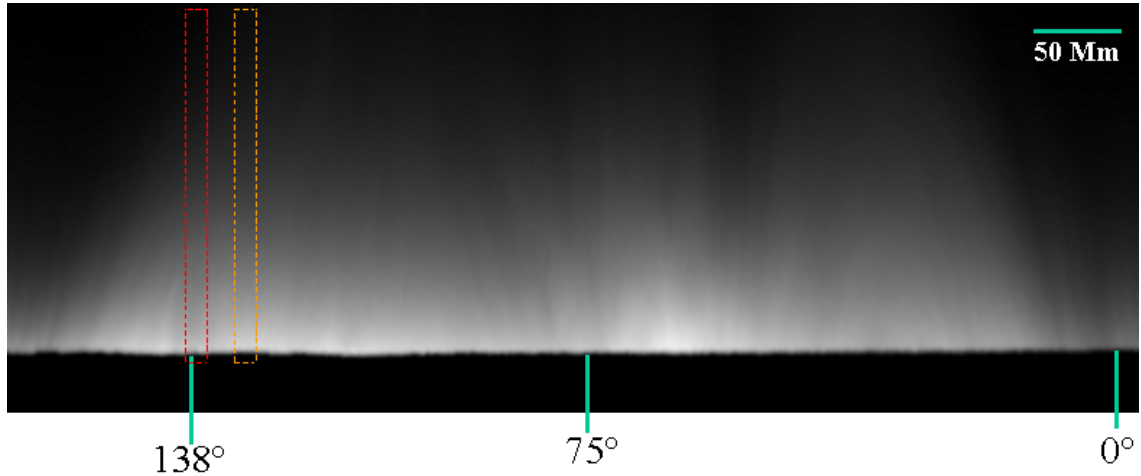

Figure 8: *W-L original image obtained during the 11$^{th}$ July 2010 total eclipse and converted into polar coordinates, showing a weaker radial gradient and the less contrasted structures of the coronal intensities. The red and orange dotted lines indicate the position of the radial cuts, along and outside the cavity.*

Figure 9 shows selected tangential cuts taken at different radial distances, between 12 Mm and 53 Mm of height, from the SWAP image taken in 174A. It shows considerably depressed intensities of this cavity region, reaching values of order of 60 %. The deficiency ratio seems to be constant for heights up to 30 Mm. Above 35 Mm, the cavity effect seems to completely disappear.

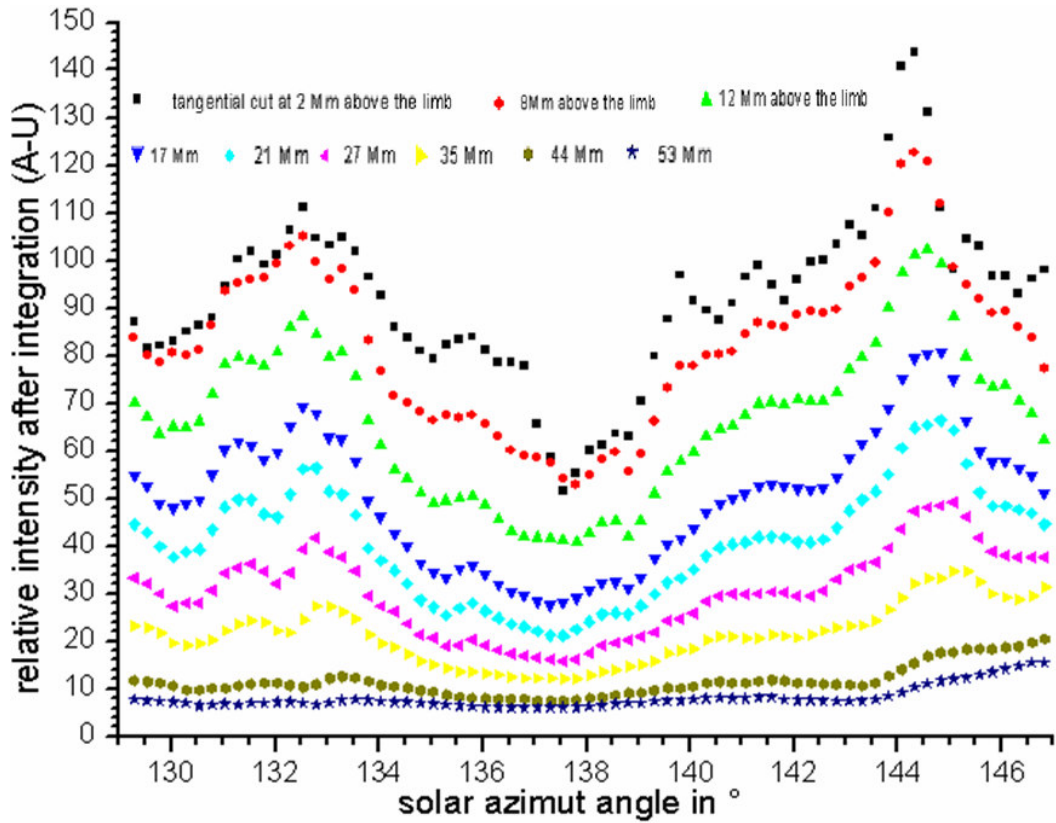

Figure 9: *Tangential cuts taken from 9 to 53 Mm around the SWAP cavity region showing the amplitude variations of the depression with heights counted from the lunar edge.*

The extension of the cavity seen with SWAP needs more analysis in order to evaluate the deficiency in the Fe IX/X lines and to interpret it, using emission measurements, assuming a reasonable value of the temperature inside the cavity. In addition, line of sight effects should be discussed. The graph of Figure 10, showing the radial variations, is obtained by integrating over a 20Mm width (see Figure 7). It shows the intensity profiles taken along the cavity and outside the cavity.

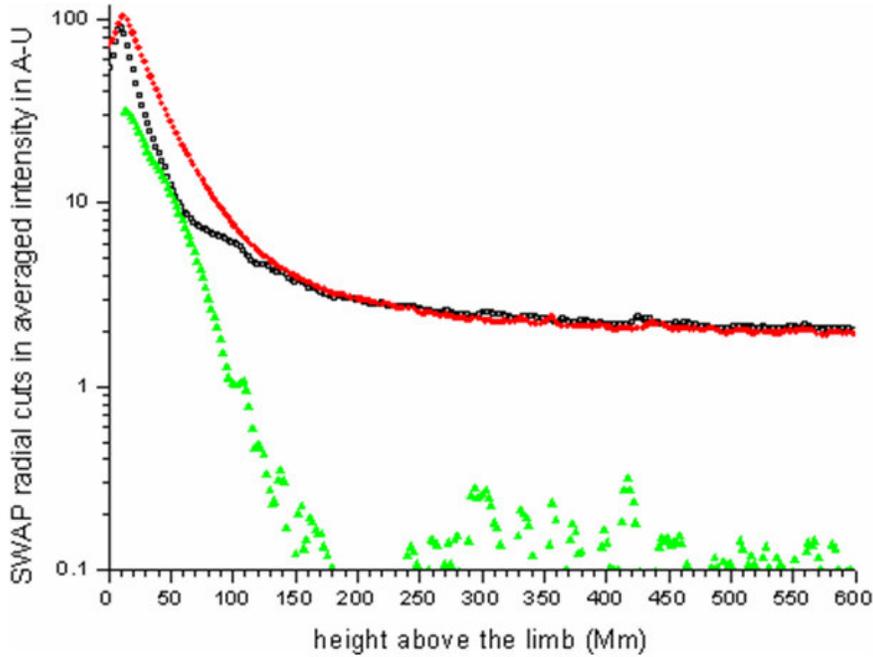

- along the cavity
- outside the cavity
- difference outside - along the cavity

Figure 10: *Radial intensity profiles from SWAP taken in the region of the South East cavity at 138° and difference between the radial cuts inside and outside the cavity, in Log scale and using a 20 Mm width of integration. The red curve corresponds to the radial cut, taken outside the cavity, the black one corresponds to the profile through the cavity and the green curve is the absolute difference between radial cuts taken along and outside the cavity. The deficiency reaches a relative maximum of about 55 %.*

Providing the temperature inside the cavity is near the temperature Ti given by the ionisation balance (.6 to 1 MK temperature)**,** the emission in the 174A Fe IX/FeX emission lines is proportional to the coronal electronic density squared ($Ne^2$) when the excitation of the line is collisionally dominated (see Allen, 1975, Thomas, 2003 and also the classical monographs by Shklovskii, 1965 or by Billings, 1966). In SWAP 174A, we find a contrast ratio at the height of 40 Mm of the order of 0.4, see Figures 9 and 10. The effective length of emissivity $L_{eff}$, observed along the line of sight, is different for SWAP 174A emissions and for W-L intensities due to the Thomson scattering from electrons, which is only linearly proportional to Ne. In the case of 174A, intensities are proportional to $Ne^2.f(Ti)$ and accordingly, the effective length of the line of sight integration $L_{eff}$ should be smaller **(Allen, 1975).** However, in the case of 174A the temperature variation along the line of sight is another factor that should be taken into account. In W-L the temperature shapes the corona only in the radial direction, **assuming the hydrostatic equilibrium**, but usually this temperature is considered constant over several hydrostatic scale heights (see for ex. November and Koutchmy, 1996). The radial gradient of Figure 10 are in agreement with the classical values (Shklovskii, 1965; Billings 1966) showing a decrease by a factor 10 in intensity between 1 to 1.3 solar radii. In the comparison between W-L and EUV images, we even found some similarity with the emission of the Fe XV line at 284A, as seen in Figure 7 page 172 reported by Allen, 1975, where a difference of the electron density from W-L (Thomson scattering, only) at eclipse condition and from the Fe XV emission line at 284A (which corresponds to rather higher temperatures) is reported. In order to explain the discrepancy, Allen, 1975 introduced an irregularity factor with a value 4 near the surface, but his study concerns the general corona with a spatial resolution significantly lower than the today resolution where structures are now analysed.

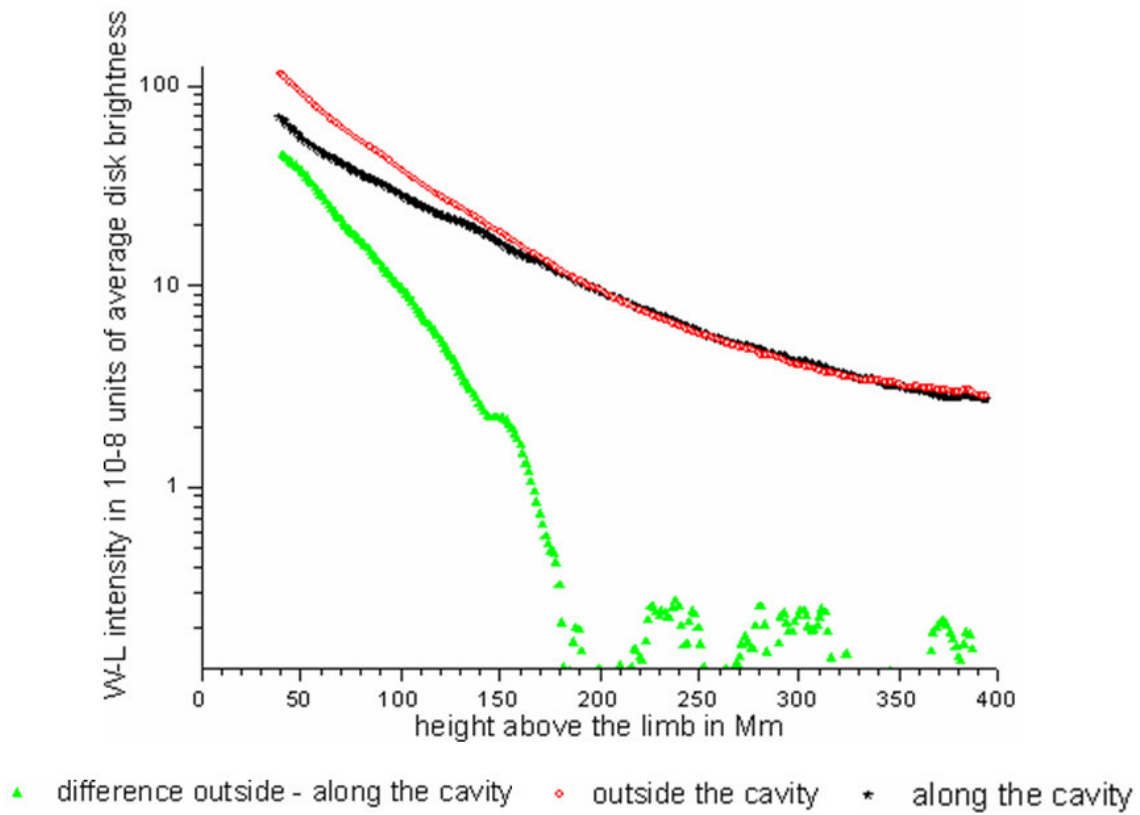

Figure 11: *Radial cuts along the cavity in red, outside in dark and green is the absolute difference along and outside the cavity, from the W-L images. The contribution of the F-corona was subtracted before making the cuts. Note that* the heights of *the continuum measured here is well above, in the radial direction, the height of the continuum reported from the flash spectra.*

In the case of a homogeneous W-L corona and for these radial distances (near the 40 Mm heights), from November and Koutchmy (1996) the effective length of integration $L_{eff}$ is of order of 1 solar radius or 700 Mm. From the value of the intensity deficit observed in W-L (Figure 11), and, assuming a negligible value of the electron density inside the cavity, it is possible to evaluate the effective "extension" of the cavity along the line of sight, close to the horizontal direction. We assume that the cavity is extended along the line of sight during the day of the eclipse. Figure 8 shows the W-L azimuthal intensities in solar heliocentric angles in Log scale intensity as a function of the solar radial distance. From the previous Fig. 11, at 40 Mm heights we measure in W-L a ratio of 1/4, which is typically 2 times lower than what we measured in 174A with SWAP. In W-L, the extension of the cavity is then evaluated to be 700/4 = 175 Mm along the line of sight, assuming it is completely empty.

We also used images from the EIT (Extreme-ultraviolet Imaging Telescope) of SOHO (SOlar Heliospheric Observatory) from NASA and ESA, launched in December, 2$^{nd}$ 1995. This instrument provides images of the hot corona (1.5 MK) in 195A FeXII iron and in the 304A HeII lower temperature lines. The field of view is 1024x1024 with a 2.6 arcseconds pixels (see EIT of SOHO web site http://umbra.nascom.nasa.gov and http://sohowww.nascom.nasa.gov). Images in 195A although noisier, allowed us to examine the cavity at higher temperatures. Figure 12 shows the 12 stacked fits images of the EIT 195A converted in polar coordinates.

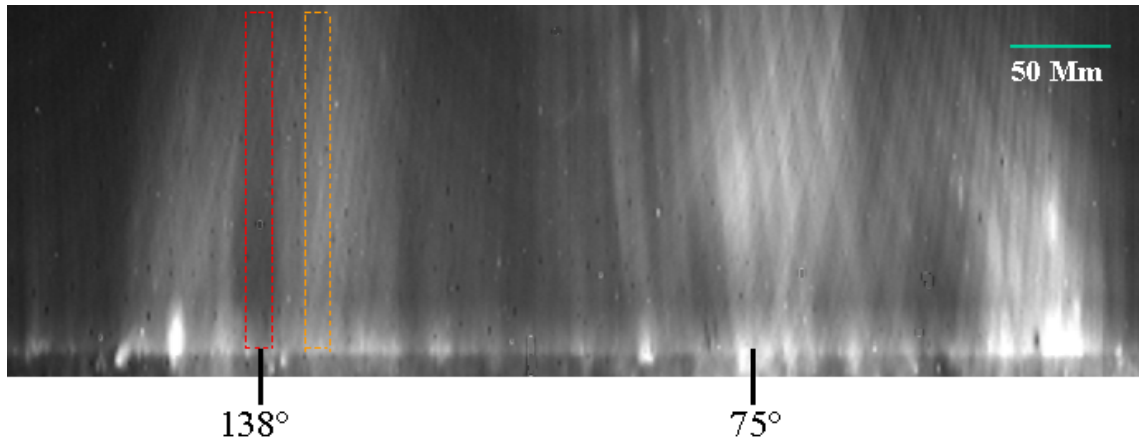

Figure 12: *12 stacked EIT 195 images to show the South-East cavity at 138° between 17h36 to 20h36 mn UT, and the emissivity of the surrounding corona*

The deficiency values at 195A are surprisingly found to be only about 6.6 %, without taking into account the effect of the spurious scattered light (stray light effect). In this "hotter" emission line of about 1.5 MK the cavity contrast seems reduced, a result similar to the one recently described by Habbal et al. (2010 and 2011) and by Pasachoff et al. (2011) for eclipse cavities. Moreover in Fig. 12, in the radial direction, we see in 195 a higher "scale height" or more precisely, a weaker radial gradient than what was seen in 174A, see Figure 7, providing that we trust the intensity variations given in the EIT (SoHO) filtergrams. This effect is also present in W-L, see the radial scans shown in Figure 11. A weaker W-L radial gradient corresponds to a higher hydrostatic temperature just because the W-L intensities reflect the values of the electron densities.

III) Extension of the cavity region

We used 20 stacked SWAP images to increase the signal to noise ratio, and extracted the cavity region corresponding to prominence 2, and we repeated this processing every 12 hours between July, 9th to July 13th, for evaluating the cavity extension, see figure 13.

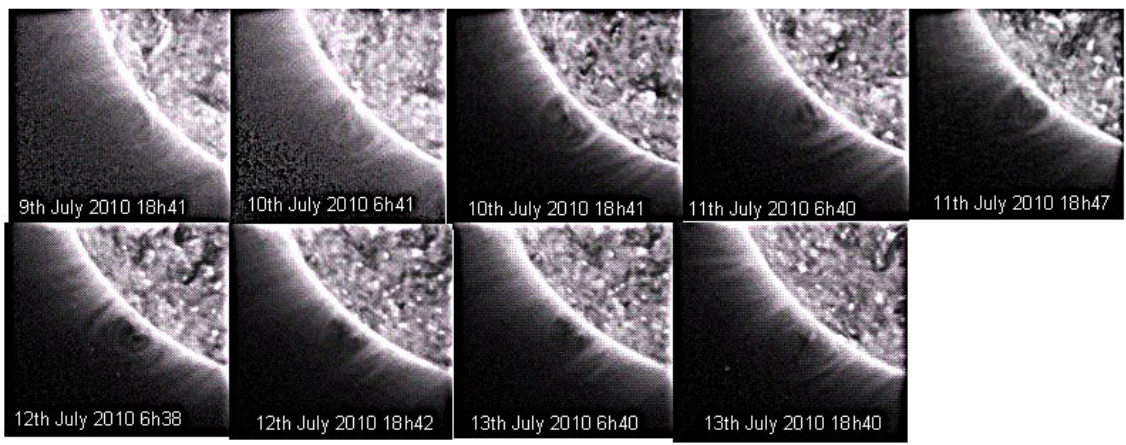

Figure 13- The cavity region of prominence 2 seen every 12 hours of time intervals using SWAP images, from July 9th to July 13th. The cavity can be better identified over the limb from July $10^{th}$ near 18h to July $13^{th}$ near 6h e.g. during 2.5 days. Plasma flows inside the cavity core are also present.

Although the exact geometry of the cavity is rather questioning, it seems that the limb passage is occuring 1/2 day after the eclipse observation.

IV) Discussion and conclusion

The 2D mapping of the HeI/HeII relative intensity ratio shows that the maximum value (in agreement with Hirayama) is about 3.3 in the central parts of the prominence and, interestingly, that the ratio decreases abruptly by a factor 2 over 10 Mm going to the outer parts, illustrating the behavior through the region where the temperature, and consequently the heating rate, is drastically increasing. A quite similar result, based on a diagnostics which uses completely different lines, was deduced in Stellmacher et al. 2003.

At larger scales, the photometric cuts show the typical depressed coronal intensities, in both 174A and W-L, revealing the cavity region around the prominence with a deficit of density.

From the details of the summed flash spectra revealing the behavior of the continuum where the prominence is recorded in He lines, we obtain evidence of a rather low electron summed density inside the prominence, not significantly larger than the one that is missing in the surrounding electron corona at the chromospheric level. This is a result of using the true continuum levels (Thomson scattering by free electrons), thanks to the use of the spectral data and it needs to be confirmed at the forthcoming eclipses.

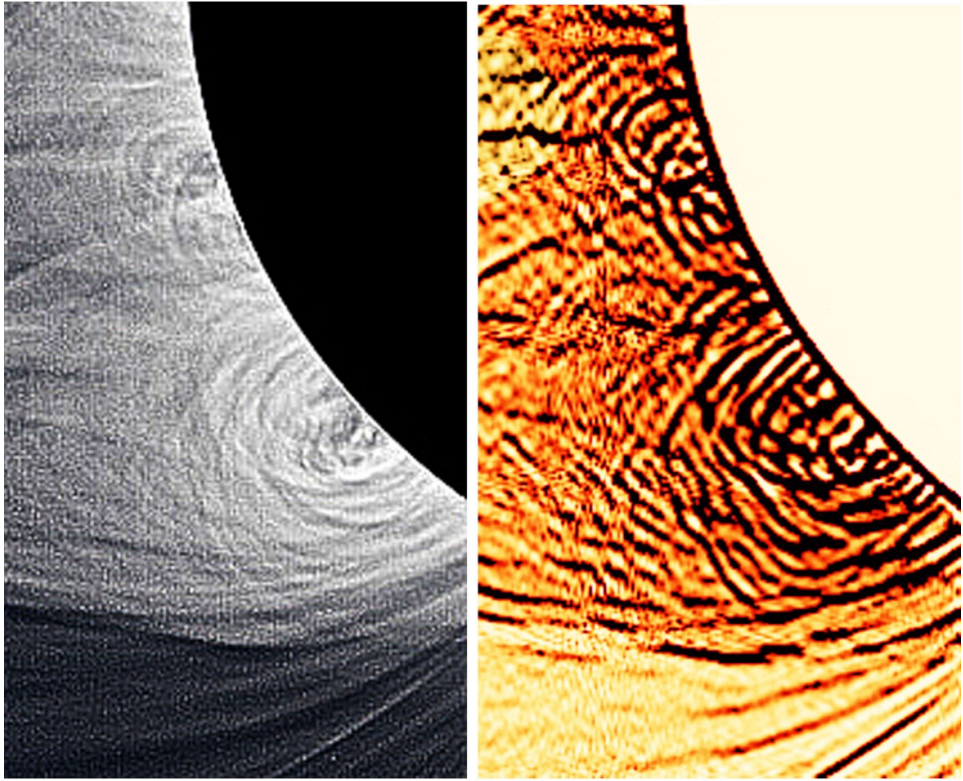

Figure 14- *Extract from the highly processed W-L eclipse image made for showing the contour of loops and fine scale details after removing the larger scale gradients (low spatial frequencies filtering). At left extract from the best M. Druckmuller image (see Habbal et al. 2011) and at right, in negative, the same image after processing using the Madmax operator (see Tavabi and Koutchmy, 2012). Note that the processing removes the photometric properties of the picture and the cavity intensity deficit is completely washed out.*

Furthermore, the extension of the cavity along the line of sight was evaluated, without taking into account heterogeneity. It is of order of 175 Mm in W-L and possibly different in the SWAP 174A emissions (.6 to 1 MK temperature). The evaluated extension of the cavity could also depend of the temperature. Harvey (2001) reported that cavities are hot plasma inside the filament channels. An extensive discussion of the temperature effects in cavity regions was recently published by Habbal et al (2010, 2011); our results tend to confirm their conclusions. In principle, the extension of the cavity channel region can be evaluated, assuming no temporal change, thanks to rotation effects, see figure 13. We also evaluated several movies made from AIA 171 images processed to remove the radial gradient in order to better see dynamical effects and, eventually, guess the geometry of the cavity using different points of view offered by the rotation effect. They show a lot of small and large scale dynamical effects, in addition to showing that the cavity is indeed surrounded by many tiny irregular and moving loops that are difficult to measure. The analysis of dynamical effects is beyond the scope of this paper and here we just mention the existence of the effect in the frame of a search for the 3D structure of the cavity. The overall impression is that the cavity region, at least for the case of prominence 2, is "active", that its extension is however limited, see also figure 13, and that its geometrical extension is not very much different from what is obtained from the evaluation made using the W-L diagnostics (175 Mm), without taking into account irregularities. Irregularities play an important role in case of coronal emission lines, because their $Ne^2$ dependence and, accordingly, the geometrical extension of the cavity can be longer than the effective emission length of a coronal line. In addition, temperature effects can play a role because the emission measure of each line depends of the ionization temperature of the corresponding emitting ion.

The cavity is more elongated along the local solar parallel than it is along the local longitudes, see figure 13. Assuming the duration of the limb transit of the cavity is 2.5 days and taking into account its low latitude and a rotation rate of 28 days, its geometric extension is of order of 240 Mm, not far from the extension deduced from the W-L photometry.

A highly processed and magnified W-L image, see figure 14, illustrates the overlapping effects along the line of sight, showing several arches (called loops in case of coronal EUV filtergrams) seen in the region of prominence 2. It confirms the importance of irregularities and illustrates the difficulty of evaluating the extension in latitudes.

In conclusion, the comparison between deep SWAP and W-L eclipse images is very enlightening and permits a more extended discussion of the cavity phenomenon. We plan to repeat these experiments and observations during the next total solar eclipse that will occur in Australia on November 13$^{th}$ 2012 and use again the SWAP data, simultaneously with the higher spatial resolution AIA images. The Sun in November 2012 should be more active than it was for July 2010, and we expect new results.

**Acknowledgements**

We first address many thanks to the SWAP team for the 174 images, which were provided by the Proba 2 Belgium consortium and we specially thank David Berghmans and Anik De Groof for their help in organizing the collaborative GI programs; the whole ROB Proba 2 team should be congratulated for succeeding in obtaining very good SWAP sequences during the 2010 total solar eclipse.

We warmly thank the AIA teams (SDO mission) for providing EUV high resolution images from space that we used in this paper, courtesy of NASA, see AIA/SDO http://sdo.gsfs.nasa.gov and AIA Instrument http://aia.lmsal.com ; we thank the US laboratories involved in the development of this wonderful experiment, including the SAO (Cambridge) and the Lockheed group near Stanford, Ca.

We also thank the EIT SoHO team for still providing good images after 14 Years of operation, courtesy of both ESA and NASA as a result of very successful collaborations.

We thank Zadig Mouradian, Jean-Claude Vial, G. Stellmacher, E. Wiehr, Frederic Auchere, and Philippe Lamy for discussions during the genesis of this paper and later, and Eleni Dara and Leon Golub for reviewing it. Antoine Llebaria helped us with the IDL program to convert images into polar coordinates, M. Druckmüller provided an excellent processed eclipse image from Tatakoto and Jean Mouette successfully took the white light images during the total solar eclipse of 11th July 2010 in French Polynesia and processed them. Observations were supported by CNES (France), in the framework of an action to complement the data collected by the Picard space mission.

Finally, we sincerely thank our Referee for helping us in deeply improving the paper.

**References**

Allen, C.W. 1975, MNRAS, **172**, 159

Bazin, C. Koutchmy, S. Tavabi, E. 2011 "The He I and He II chromospheric shells and the transition region" IAGA II, 4-8 December 2009, Cairo Egypt, in proceedings edited by Hamed Hady and Luc Dame, Solar Wind and space environnement interaction.

Billings, D. A. 1966, Guide to the Solar Corona, Academic Press New-York, London

Berghmans, D. Hochedez, J-F. Defise, J.M. Lécat, J.H. Nicula, B. Slemzin, V. et al.; 2006, Adv. Space Research, **38**, 1807.

De Groof A., Berghmans D., Nicula B., et al. 2008**a**, Solar Phys. **249**, 147-163

De Groof, A. Berghmans, D. Defise, J.M. Nicula, B. and Schuehle, U. 2008**b**, 12th European Solar Physics Meeting, Freiburg, Germany, online at http://espm.kis.uni-freiburg.de/, p.2.116

Defise J-M. et al. 2007, Proceedings of the SPIE, vol. **6689**, pp. 66890S-12

Engvold, O. Hirayama, T. Leroy, J.L. Priest, E.R. and Tandberg-Hanssen, E. 1990, in I.A.U. Coll. 117. Dynamics of Quiescent Prominences, ed. by V. Ruzdjak, E. Tandberg-Hanssen. Lecture Notes in Physics, Berlin Springer Verlag, vol. 363, 1990, p.294

Filippov, B. Koutchmy, S. and Tavabi, E. 2011, Solar Phys. Topical Issue on Proba 2 (in press)

Habbal, S. et al 2010, Astrophys. J. **719**, 1362

Habbal, S. et al. 2011, Astrophys. J. **734**, 120

Halain, J-P. Berghmans, D. Defise, J-M. Renotte, E. et al. 2010, in : Arnaud, M. Murray, S. Takahashi, T. (Eds.) Space Telescopes and Instrumentation 2010: UV to Gamma Ray, Proc. SPIE **7732**, 77320P.

Harvey, K. 2001, "Coronal Cavities" in Encyclopedia of Astronomy and Astrophysics, Murdin, P. Ed., Nature Publishing Group, Hampshire UK and Institute of Physics Publishing 2001 Bristol UK

Hirayama, T. 1964, Publ. Astron. Society of Japan **16**, 104

Hirayama, T. and Nakagomi, Y. 1974, Publ. Astron. Soc. Japan, **26**, 53

Hirayama, T. and Irie, M. 1984, Solar Phys. **90**, 291

Hirayama, T. 1971, Solar Phys. **17**, 1, 50

Jejcic, S. and Heinzel, P. 2009, Solar Phys. **254**, 1, 89

Koutchmy, S. Lebecq, C. and Stellmacher, G. 1983, Astron. Astrophys. **119**, 261

Koutchmy, S. and Lamy, P. 1985, "Properties and interactions of interplanetary dust"; Proceedings of the Eighty-fifth IAU Colloquium, Marseille, France, July 9-12, 1984 (A86-42326 20-90). Dordrecht, D. Reidel Publishing Co., 1985, p. 63-74.

Koutchmy, S. Filippov, B. and Lamy, Ph. 2007, in "The Physics of Chromospheric Plasmas", ASP Conference Series, Vol. 368, Proceedings of the conference held 9-13 October, 2006 at the University of Coimbra in Coimbra, Portugal. Edited by P. Heinzel, I. Dorotovič, and R. J. Rutten. San Francisco: Astronomical Society of the Pacific, p.331

Koutchmy, S. Bazin, C. Druckmuller, M. et al. 2011, Astrophys. J. (submitted)

Kubota, J. and Leroy, J-L. 1970, Astron. Astrophys. **6**, 275

Labrosse, N. Heinzel, P. Vial, J-C. Kucera, T. Parenti, S. Guar, S. Schmieder, B. and Kilper, G. 2010 Space Sci. Rev. **151**, 243

November, L. and Koutchmy, S. 1996, Astrophys. J. **466**, 512

Pasachoff, J. et al. 2011, Astrophys. J. **734**:114

Saito, K. and Tandberg-Hanssen, E. 1973, Solar Phys. **31**, 1,105

Saito, K. and Hyder, C. 1968 Solar Phys. **5**, 1, 61

Shklovskii, I. S. 1965, in Physics of the Solar Corona, 2$^{nd}$ Edition Vol. 6, Pergamon Press, Oxford-London-Edinburgh, New-York, Paris-Frankfurt, p 249-251

Sirk, M.M. Hurwitz, M. and Marchant, W. 2010, Solar Phys. **264**, 2, 287

Stellmacher, G. Wiehr, E. and Dammasch, I.E. 2003, Solar Phys. **217**, 133

Tandberg-Hanssen, E. 1995, in The Nature of Solar Prominences, Kluwer Acad. Pub. Dordrecht, The Netherlands, 308 p.

Tavabi, E., Koutchmy, and S. Ajabshirizadeh, A. 2012, Solar Phys. (in press in Topical Issue on "Solar Image Processing in the Petabyte Era"), preprint in **arXiv:1104.5580**

Thomas, R.J. 2003, in Solar Polarization 3, ASP Conf. Series, Vol. **307**, J. Trujillo Bueno and J. Sanchez Almeida, eds. p. 497

**Appendix: Schematic of the flash spectra experiment used in Hao French Polynesia during the 11$^{th}$ July 2010 total eclipse.**

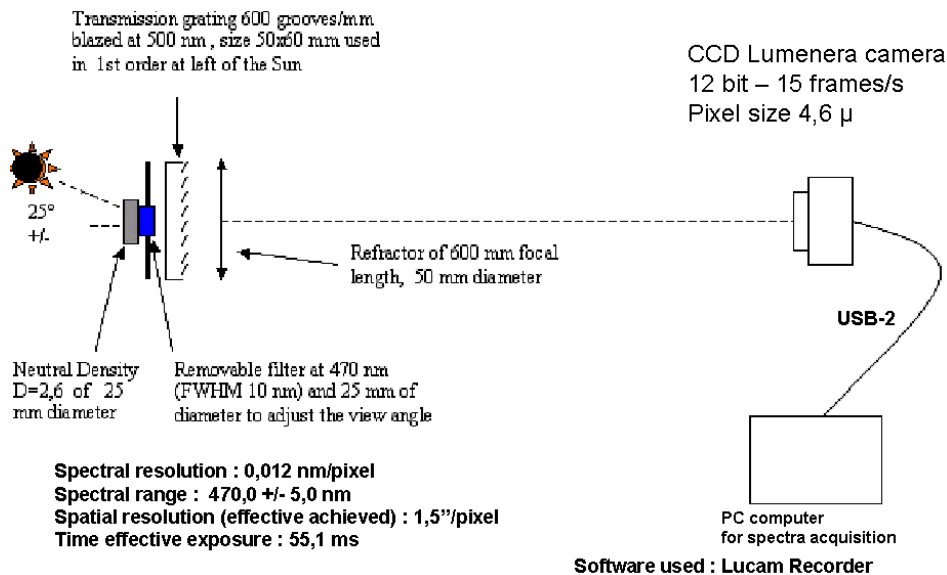

**The setup, consisting of a transmission diffraction grating (used in the first order) placed in front of the refracting achromatic lens, is fixed on an equatorial mount for guiding. The angle of ~ 25° indicates the solar spectrum deviation with the optical axis of the refractor.**